\documentclass[conference]{IEEEtran}
\usepackage{tcolorbox}
\usepackage{wrapfig}
\usepackage{breakurl}
\usepackage{bold-extra}
\usepackage{lipsum}
\usepackage{amsfonts}
\usepackage{graphicx}
\usepackage{stfloats}
\usepackage{enumitem}
\usepackage{xspace}
\usepackage{comment}
\usepackage{caption}
\usepackage{epstopdf}
\usepackage{algorithmic}
\usepackage{algorithm2e}
\usepackage{tabularx}
\usepackage{setspace}
\usepackage{todonotes}
\usepackage{xspace}
\usepackage{xcolor}
\usepackage{wrapfig}
\usepackage{array}
\usepackage{makecell}
\usepackage{lipsum}
\usepackage{tcolorbox}
\usepackage{tabularx,booktabs}
\newcolumntype{C}{>{\centering\arraybackslash}X} 
\usepackage{caption}

\newcolumntype{P}[1]{>{\centering\arraybackslash}p{#1}}
\usepackage[flushleft]{threeparttable}
\usepackage{array,booktabs,makecell}
\usepackage{booktabs}
\usepackage{dcolumn,tipa}
\newcolumntype{d}[1]{D{.}{.}{#1}} 
\usepackage[usestackEOL]{stackengine}
\usepackage{multirow}

\usepackage{tabulary}
\usepackage[normalem]{ulem} 
\usepackage{multirow,array}
\usepackage{hyperref}   
\usepackage{pdflscape}  
\usepackage{colortbl}

\definecolor{light-gray}{gray}{0.96}
\definecolor{LightCyan}{rgb}{0.88,1,1}

\definecolor{light-gray}{gray}{0.96}

\newcommand{\defn}[1]{\textbf{\emph{#1}}}

\newif\ifcomments
\commentstrue

\begin{document}

\title{A Policy Driven AI-Assisted PoW Framework}

\author{\IEEEauthorblockN{Trisha Chakraborty\IEEEauthorrefmark{1},
Shaswata Mitra\IEEEauthorrefmark{2}, Sudip Mittal\IEEEauthorrefmark{3}, Maxwell Young\IEEEauthorrefmark{4}}
\IEEEauthorblockA{Department of Computer Science \& Engineering, Mississippi State University\\
\IEEEauthorrefmark{1}tc2006@msstate.edu,
\IEEEauthorrefmark{2}sm3843@msstate.edu,
\IEEEauthorrefmark{3}mittal@cse.msstate.edu,
\IEEEauthorrefmark{4}myoung@cse.msstate.edu\\}}
\maketitle


\begin{abstract}

Proof of Work (PoW) based cyberdefense systems require incoming network requests to expend effort solving an arbitrary mathematical puzzle. Current state of the art is unable to differentiate between trustworthy and untrustworthy connections, requiring all to solve complex puzzles. In this paper, we introduce an Artificial Intelligence (AI)-assisted PoW framework that utilizes IP traffic based features to inform an `adaptive' issuer which can then generate puzzles with varying hardness. The modular framework uses these capabilities to ensure that untrustworthy clients solve harder puzzles thereby incurring longer latency than authentic requests to receive a response from the server. Our preliminary findings reveal our approach effectively throttles untrustworthy traffic.

\end{abstract}


\section{Introduction}
\label{sec:Introduction}

A distributed denial-of-service (DDoS) attack is a malicious attempt to disrupt the normal traffic of a targeted server, service, or network by overwhelming the target or its surrounding infrastructure with a flood of Internet traffic. A possible defensive strategy is an effective Proof-of-Work (PoW) based system ~\cite{nakamoto2008bitcoin, Gupta2018ProofOW,le2012kapow}. A PoW system works by requiring incoming network requests to expend effort solving an arbitrary mathematical puzzle to prevent anybody from attacking the system. In PoW based systems, a client has to commit some computation (CPU cycle, bandwidth, etc.) to solve a puzzle to prove her authenticity.

PoW systems generally consist of three parts: \textit{issuer}, \textit{solver}, and \textit{verifier}. The issuer (also called a generator) issues the puzzle to the solver, which solves them and sends the solution to the verifier. In a simple networked client-server environment, the server contains the issuer/generator and the verifier components, and the client is the solver.

In this paper, we build an Artificial Intelligence (AI)-assisted PoW framework. We create an `adaptive' issuer which can generate puzzles with varying hardness. The idea behind the system is to penalize untrustworthy connections by issuing them `hard' puzzles and at the same time give `easy' puzzles to trustworthy requests. \textit{Critically, these challenges introduce latency in the environment for untrustworthy connections}. The distinction between trustworthy/untrustworthy clients can be made by using the incoming traffic specific features. In other words, an AI subsystem can compute a \textit{reputation score} for an incoming request, that can guide the puzzle generator.


Our framework has two useful properties. First, each client pays a cost for utilizing the system, and this cost increases as the client's reputation score worsens. 
Second, the amount of work inflicted by a puzzle is {\it adaptive} and can be tuned. The framework will ensure that clients with bad reputation scores incur longer latency to receive a response from the server than authentic requests. This latency is beneficial to the network under attack to slow down the incoming malicious traffic.


By creating this AI-assisted PoW puzzling scheme with the above mentioned properties, we strengthen the defensive posture of organizations that use these strategies. Our framework is modular and each component can be customized. Components include: an AI model that generates a reputation score, a puzzle generator, a puzzle solver, a puzzle verifier, and a policy that maps a reputation score to puzzle hardness. Regarding the last component, a network administrator may specify a policy based on her specific security needs. Next, we describe our system architecture and evaluations. Our results demonstrate that the AI assisted PoW puzzle introduces latency in the network for clients with bad reputation scores. 



\begin{figure}[h]
\includegraphics[width=9.3cm]{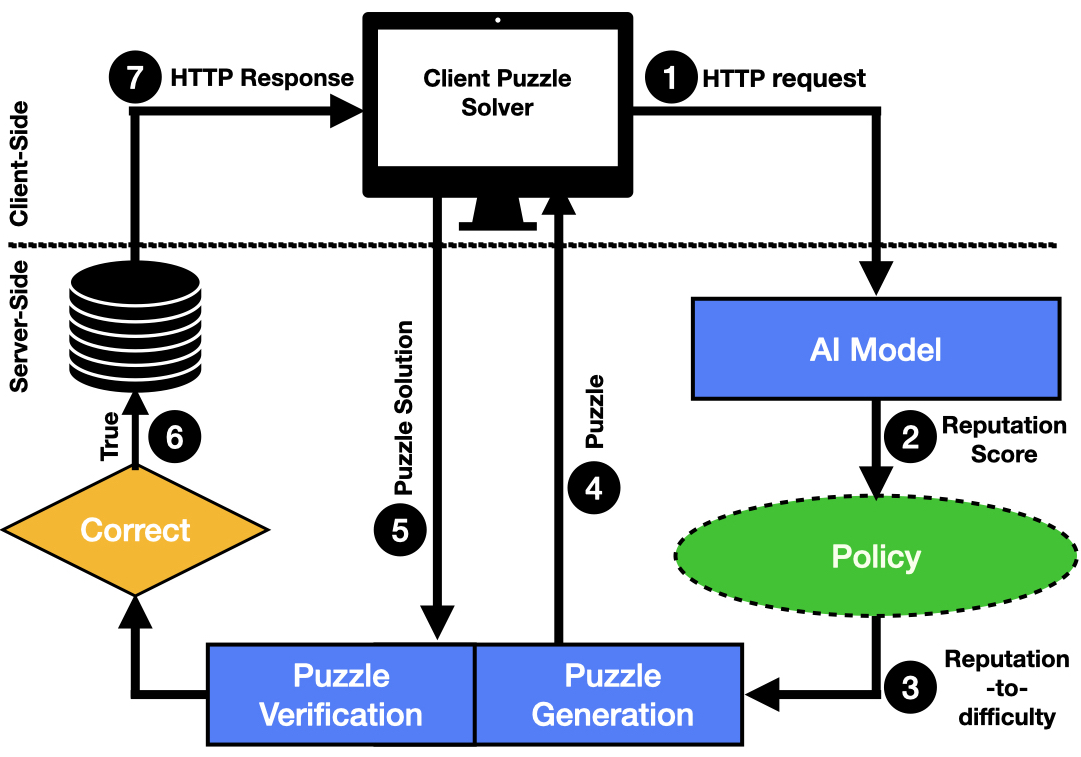}
\caption[]{An illustration of our framework. (1) The client issues an HTTP request to the server. (2) The AI model inspects the features of the request as input and produces a reputation score. (3) The policy module maps the reputation score to a difficulty. (4) The puzzle generation module generates a puzzle and sends it to the client to solve. (5) The puzzle verification module verifies the puzzle solution returned by the client. (6) On solving the puzzle correctly, the server is informed. (7) The server responds with the requested resource.}
\label{fig:Architecture}
\vspace{-5mm}
\end{figure}
 
\section{System Architecture}
The architecture of our AI-assisted PoW framework is depicted in Figure \ref{fig:Architecture}. The components of this framework are described below.




\subsubsection{AI Model} To create a proof of concept we use DAbR \cite{renjan2018dabr} as our AI model. DAbR is an euclidean distance-based technique that generates a reputation score for an IP address by learning from previously known malicious IP addresses and their attributes. 
The model generates a reputation score for an IP with an accuracy of 80\%. The reputation score is normalized to a scale of 0 - 10, where a higher score represents a more untrustworthy client. This reputation score serves as an input to the policy module as a variable mapped to puzzle hardness.


\subsubsection{Policy} The policy module takes as an input a reputation score $R$ (range $[0,10]$). Intuitively, a high reputation score of $10$, denoting an untrustworthy client should be allotted a harder puzzle than a relatively better reputation score of $9$. Hence, proper tuning of the difficulty is desired for fine-grained reputation scores. A \textit{policy} is a rule-based strategy for mapping the reputation score of a client to the appropriate puzzle difficulty. In Section \ref{sec:evaluation}, we implement and analyze three such policies.

\subsubsection{Puzzle generation} Our puzzle generation module issues a PoW puzzle. A puzzle is generated by collecting request related data, i.e., timestamp and unique seed (for mitigating pre-computation attacks), and a difficulty value as defined by the policy module, all of which is relayed back to the client.

\subsubsection{Puzzle solver} The data received from the puzzle generation module are concatenated with the client's IP address to form a string that is not altered. To this, a $32$-bit string is added, which the client modifies upon each hash function evaluation. The client performs evaluations on this input until it finds an output with a prefix of $d$ zeros; we refer to this as a \defn{$\mathbf{d}$-difficult puzzle}. Our strategy is to assign higher difficulty of puzzles to clients with bad reputations scores.
\subsubsection{Puzzle Verification} Puzzle verification is light weight block used to verify the clients solution and offer response if correct solution is returned.

\section{Evaluation}
\label{sec:evaluation}
For our proof of concept implementation we utilize DAbR \cite{renjan2018dabr}, which uses IP protocol based features to generate a reputation score. In this section, we discuss three policies used for a preliminary evaluation of our framework.
 
 


\subsection{\textbf{Policies $\mathbf{1}$ and $\mathbf{2}$: Linear mapping}} For this evaluation, $R$ take on values in the set- $\{0, 1,..., 10\}$. The lower the reputation score, the more confidence we have that the client is trustworthy; conversely, the higher the reputation score, the more we suspect that the client is untrustworthy. Consequently, the difficulty of a puzzle assigned to a client increases with the client's reputation score. It takes $31$ ms on average to solve a $1$-difficult puzzle, and this time increases with difficulty.


For Policy $1$, we map a $1$-difficult puzzle to a client with a reputation score $0$, a $2$-difficult puzzle to a client with a reputation score of $1$, and so on. Figure \ref{fig:policy} shows that the latency increases with the increase in reputation score. However, the latency does not grow significantly as the  reputation score increases. To address this, we evaluate Policy $2$, where the easiest puzzle has difficulty $5$. Thus, we map a $5$-difficult puzzle to the client with reputation score $0$, a $6$-difficult puzzle to a client with a reputation score of $1$, and so on. As a result, the latency increases significantly with higher reputation scores, delaying service for untrustworthy clients. 

\subsection{\textbf{Policy $\mathbf{3}$: Error range mapping}}



\begin{figure}[t]
\centering
 \includegraphics[width=8.7cm, trim = 3cm 8.3cm 3.5cm 9cm, clip]{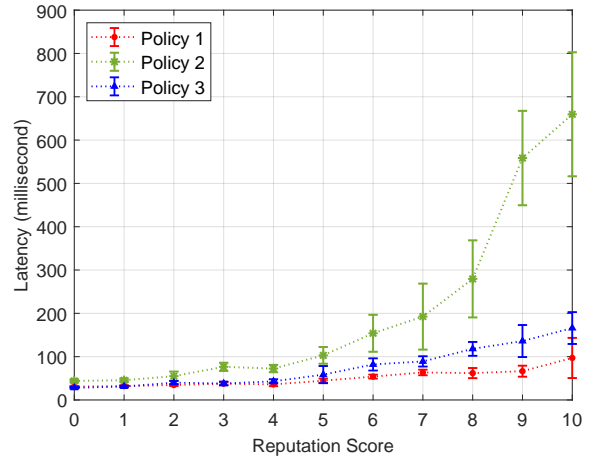}
\caption[]{An evaluation of our three implemented policies. The median of 30 trials is reported for each reputation score.}
\label{fig:policy}
\vspace{-5mm}
\end{figure}

In this policy, we consider the error $\epsilon$ from DAbR system \cite{renjan2018dabr}. Note that, given this error, the resulting IP reputation score might be higher or lower than the ground truth. Our Policy 3 attempts to correct for this in the following way. All reputation scores $s_i$ are in the interval $[0, 10]$. For a score $s_i$, the difficulty value is a value chosen at random in the interval 
$[\lceil d_i -\epsilon \rceil, \lceil d_i + \epsilon\rceil ]$, where $d_i = \lceil s_i+1\rceil$.
Figure \ref{fig:policy} shows how the rate of increase in the latency for Policy $3$ is between our two previous policies.\\

\noindent{\bf Summary.} Our evaluation illustrates how the amount of latency can be tuned given different mappings between reputation scores and puzzle difficulty. Thus, our framework offers flexibility and can accommodate the specific security demands for a range of network settings.

\section{Conclusion}
In this paper, we proposed and evaluated an AI-assisted PoW framework. It employs an AI model that uses IP-based features to output a reputation score. This score serves as an input to an `adaptive' puzzle generator which creates puzzles of varying difficulty. In this way, the framework imposes higher latency on untrustworthy clients. We evaluated our proposed framework, showcasing how it can help defend organizations against networking threats.

\bibliographystyle{plain} 
\bibliography{refs}

\end{document}